\documentclass[aip,reprint]{revtex4-1}
\usepackage{graphicx}
\usepackage{amsmath}
\usepackage{xcolor}
\usepackage[english]{babel}
\begin{document}

    %\preprint{}

    \title{A combined statistical mechanical and \textit{ab initio} approach to understanding H$_2$O/CO$_2$ co-adsorption in mmen-Mg$_2$(dobpdc)} %Title of paper



    \author{Jonathan R. Owens}
    \email[]{jon.r.owens@gevernova.com}
    \affiliation{Material Chemistry and Physics Lab, GE Vernova Advanced Research Center, Niskayuna, NY 12302}

    \author{Bojun Feng}
    \affiliation{AI, Software, and Robotics Lab, GE Vernova Advanced Research Center, Niskayuna, NY 12302}

    \author{Jie Liu}
    \affiliation{Material Chemistry and Physics Lab, GE Vernova Advanced Research Center, Niskayuna, NY 12302}

    \author{David Moore}
    \affiliation{Decarbonization Lab, GE Vernova Advanced Research Center, Niskayuna, NY 12302}


    \date{\today}

    \begin{abstract}
        We study the effects of H$_2$O on CO$_2$ adsorption in an amine-appended variant of the metal-organic framework Mg$_2$(dobpdc), which is known to exhibit chaining behavior that presents in a step-shaped adsorption isotherm. We first show how the presence of different levels of local H$_2$O affects this chaining behavior and the energetics of CO$_2$ adsorption, based on a series of \textit{ab initio} calculations, giving insight into the atomic-scale environment. In particular, we predict a novel adsorbed configuration, in which H$_2$O and CO$_2$ intertwine to make a braided chain down the MOF pore. We then show how an existing lattice model can be adapted to incorporate the effect of water, and predict the CO$_2$ isotherms for the various water levels, observing a sharp shift the uptake at low partial pressures. The manifestation of this braided chain in the lattice model points to the potential emergence of a shift from cooperative capture to that of a phase transition. In addition to the physical insights, this work may serve as a launching off point for further work on this and related materials.
    \end{abstract}

    \maketitle %\maketitle must follow title, authors, abstract and \pacs

    \section{Introduction}

    Capturing CO$_2$ from industrial flue streams and the atmosphere holds promise in mitigating climate change. Much academic, governmental, and industrial effort has gone into developing materials and systems that can capture CO$_2$ on a global scale. In most practical applications, other molecular species, especially H$_2$O, will be present in the gas streams from which CO$_2$ will be captured. A systematic understanding of water's effect on CO$_2$ capture is elusive and difficult\cite{kolle_understanding_2021}. The nature of the interaction is variable, and depends on the type of CO$_2$ adsorption (chemical \textit{vs.} physical) and the material itself\cite{kolle_understanding_2021}. H$_2$O can participate chemically, by forming bicarbonate with an adsorbed CO$_2$ molecule, form hydrogen bonds with carbamic acid or ammonium carbamate when adsorbed by amines, or compete physically for the same adsorption sites\cite{kolle_understanding_2021}.

\begin{figure*}
        \centering
        \includegraphics[width=\linewidth]{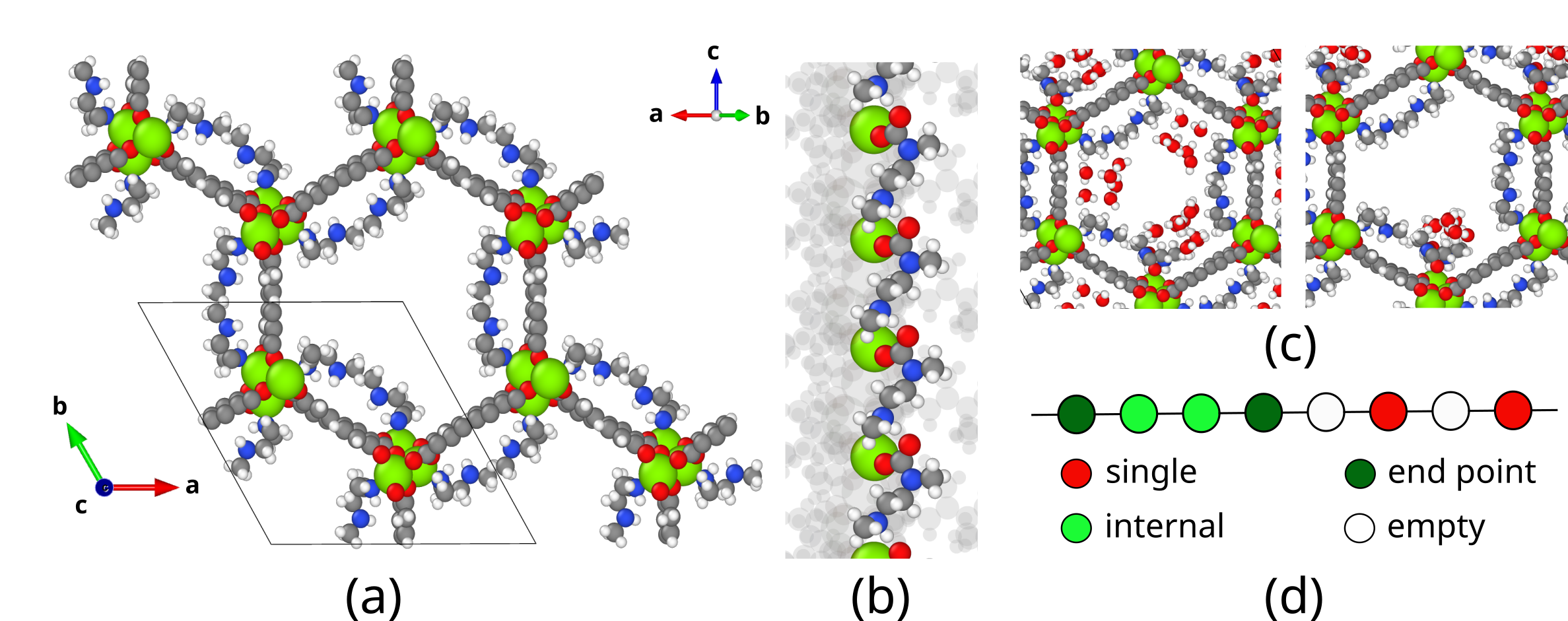}
        \caption{Overview of mmen-functionalized-Mg$_2$(dobpdc). (a) Pore-down view of the MOF with the appended amine in the absence of guest molecules. (b) Ammonium carbamate chain formation down the crystallographic \textit{c}-axis. Close inspection of the CO$_2$ shows it has passed a proton from the amine it adsorbed onto with the neighboring amine. (c) Full-pore of H$_2$O (left) \textit{vs.} a single-lane of H$_2$O (right). The single lane approximation allows us to study more localized effects. (d) Single lane lattice model, showing the possible energy states in the partition function.}
        \label{fig:struct_overview}
    \end{figure*}

    Metal-organic frameworks, consisting of metal nodes connected \textit{via} organic linkers, are crystallographically well-defined structures that been thoroughly explored for their propensity as CO$_2$ adsorbents\cite{barsoum_road_2025}. One route to enforcing CO$_2$ selectivity is the introduction of functional groups, like amines, into the frameworks. Amine-functionalized variants of Mg$_2$(dobpdc) (Figure \ref{fig:struct_overview}(a)) have been used as a high-performance sorbent for post-combustion and direct-air capture\cite{mcdonald_capture_2012,milner_diaminopropane-appended_2017,siegelman_controlling_2017,kim_cooperative_2020,mcdonald_cooperative_2015,forse_elucidating_2018,forse_new_2021}. Generally speaking, these materials form ammonium carbamate chains down the crystallographic \textit{c}-axis of the MOF, wherein the CO$_2$ inserts itself between the metal node and coordinatively bound amine (Figure \ref{fig:struct_overview}(b)), presenting with a cooperative adsorption mechanism\cite{mcdonald_cooperative_2015}. Kundu, \textit{et al.} showed that this behavior can be explained through a statistical mechanical lattice model in N-N'-dimethylethylene diamine-functionalized Mg$_2$(dobpdc) (mmen-Mg$_2$(dobpdc))\cite{kundu_cooperative_2018}. Later work identified the hysteresis responsible for this behavior\cite{edison_hysteresis_2021}. More recent work has extended this framework to account for dry\cite{marshall_cluster_2022} and humid adsorption kinetics\cite{marshall_cooperative_2024}, primarily using fits of experimental data. Compared to this previous work, ours is different in a few aspects. For one, the material we focus on is mmen-Mg$_2$(dobpdc), as opposed to the ampd-Mg$_2$(dobpdc) studied by Marshall, \textit{et al.}\cite{marshall_cooperative_2024}. Our study further is computationally-forward, interested in understanding how the local atomic environment can drive the increased CO$_2$ uptake in the presence of H$_2$O, based on \textit{structural} as well as energetic arguments. Like Kundu, \textit{et al.}, we are studying mmen-Mg$_2$(dobpdc)\cite{kundu_cooperative_2018}, but we here extend the model to incorporate the presence of H$_2$O.

    In this work, we start by exploring the energetics of the co-adsorption of CO$_2$ and H$_2$O, with varying levels of both, including detailed studies of the atomic-level structures elucidated by \textit{ab initio} calculations. We then study the physics of the co-adsorption by presenting a adapted model that accounts for the presence of H$_2$O in an adapted lattice model.

    \section{Methods} \label{sec:methods}

    \subsection{Computational methods}

    \textit{ab initio} density functional theory calculations were performed with the VASP package\cite{kresse_ab_1993, kresse_efficient_1996, kresse_efficiency_1996} at the $\Gamma$-point of the Brillouin zone owing to the large size of the supercell, with a plane wave cutoff of 600 eV and the van der Waals density function vdW-DF2\cite{klimes_chemical_2009, klimes_van_2011}, in line with previous studies on this and similar materials\cite{kundu_cooperative_2018, siegelman_water_2019, kim_cooperative_2020}. Forces were relaxed to within 0.02 eV/\AA using the residual minimization method with direct inversion in the iterative subspace (RMM-DIIS)\cite{kresse_efficient_1996, kresse_efficiency_1996, pulay_convergence_1980}. All atomic visualizations were performed with OVITO\cite{stukowski_visualization_2009}.

    To address the challenge present in H$_2$O adsorption of identifying the H$_2$O binding locations, compared to chemisorption, wherein the binding sites are well defined, we employed Grand Canonical Monte Carlo (GCMC) simulations\cite{frenkel_understanding_2023} using the RASPA2 software\cite{dubbeldam_raspa_2016} to incrementally add water molecules, one at a time, to the system. Water molecules were modeled with the TIP4P model\cite{jorgensen_comparison_1983}. The mmen-functionalized-Mg$_2$(dobpdc) was modeled as rigid with a combination of Lennard-Jones (LJ) parameters from UFF\cite{rappe_uff_1992}. The Lorentz–Berthelot mixing rules were utilized to define the Lennard-Jones (LJ) interaction parameters for unlike interactions in the system. The interactions were truncated at a spherical cutoff distance of 12.8 Å, beyond which they were not considered, and no analytical tail corrections were applied to account for the neglected interactions beyond this cutoff. For electrostatic interactions, the Ewald summation technique was employed. The partial charges for the framework atoms were determined using the Extended Charge Equilibration (EQEQ) method\cite{wilmer_extended_2012}. All simulations were conducted using 10,000 Monte Carlo cycles. Swap moves, which involve the insertion or deletion of particles, were executed with a probability of 50\% for each action. Additionally, translation moves were attempted with a probability of 30\%, while rotation and re-insertion moves were each tried with a probability of 10\%. Each water molecule was initialized at a low temperature to ensure it reached its energy minimum. After determining the optimal binding location for 1 water molecule, we translated it along the \textit{c}-axis to ensure a consistent water environment across all CO$_2$ binding sites in that lattice. This process was repeated to add between one, two, or three water molecules per CO$_2$ binding site, with the system's charges reinitialized after each addition.

    \subsection{Structure generation} \label{subsec:structure-gen}

    Another potential complication is long-distance effects of H$_2$O binding, as water forms extensive hydrogen bond networks. As we are trying to isolate specific adsorption sites and study the chaining mechanism, we want the local environment of the water to be uniform as we build up ammonium carbamate chains. We additionally want to avoid a full pore of water molecules that cause unpredictable behavior upon geometry relaxation. In our initial studies, we exploited the 6-fold symmetry of the Mg$_2$(dobpdc) unit cell, and added H$_2$O molecules symmetrically at each equivalent position and add an H$_2$O in the four \textit{c}-repetitions of the unit cell. Thus, when we discuss the scenario of 1 H$_2$O per diamine, this translated to 12 H$_2$O molecules in the super cell. This approached caused issues with small regions of condensation forming, and generally disrupted the local environment of CO$_2$ adsorption. Indeed, the observed binding energies in these cases were seemingly nonsensical, with values ranging from $-400$ to $400$ kJ/mol. This wide range and instability was attributed to the variable nature of the relaxed structures, with distant H$_2$O molecules having outsized influence of the local CO$_2$ adsorption environment (Fig. \ref{fig:struct_overview}(c)).

    We also acknowledge the inevitability of various water configurations at each CO$_2$ binding site, as the binding energy of physisorbed water can be similar across different sites. For instance, while it might seem intuitive for the first water molecule to be near carbamic acid if CO$_2$ is chemisorbed to form carbamic acid, this is not always feasible if the carbamic acid is too close to the Mg$_2$(dobpdc) walls, obstructing water entry. Additionally, the CO$_2$ binding configuration itself may vary with or without water, suggesting the possibility to consider multiple collaborative binding configurations or even conduct a statistical mechanics study over a range of CO$_2$-H$_2$O configurations.

    \section{Results} \label{sec:results}

    \subsection{H$_2$O binding locations from GCMC simulations}
    Analysis of the molecular positions generative \textit{via} GCMC H$_2$O initialization shows that H$_2$O molecules have certain preferences for specific locations within the structure regardless the number of chemisorbed CO$_2$ present. This behavior is illustrated in the left panel of Figure 1(c) and Figure 3. The tendency for neighboring amines to pair up results in the formation of an ``amine well'' on the opposite side. The first H$_2$O molecule at a given site consistently occupies this well, positioning itself as close as possible to the Mg-N covalent bond, regardless of the CO$_2$ content in the structure. Subsequent H$_2$O molecules, specifically the second and third at each site, tend to align closer to the tails of the amines. Notably, with the introduction of the third H$_2$O molecule, there is a tendency for the formation of a H$_2$O molecular chain along the c-axis. This chain formation tendency is absent when only two H$_2$O molecules are present per site. This phenomenon may be the origin of the distinct behaviors between the 2 H$_2$O and 3 H$_2$O cases, which are discussed in detail later in the paper.

    Another observation is that the locations of the H$_2$O molecules in the DFT-relaxed structures are not noticeably different from where they were inserted \textit{via} GCMC, indicating that this method is effective in helping identify the initial locations of H$_2$O molecules, as compared with manually-drawn H$_2$O, which tend to be at higher-energy locations, even when using ``chemical intuition'' to choose adsorption location.

    \subsection{Energetics of chain formation} \label{subsec:chain-energetics}

    Our model structure consists of 4 unit-cell repetitions of mmen-Mg$_2$(dobpdc) in the \textit{c}-direction. We compute the energetics of chain formation, starting from 1 CO$_2$ adsorbed \textit{via} the insertion mechanism, then 2, 3, and 4. For all of these cases, we consider 3 different H$_2$O levels: 1, 2, and 3 H$_2$O per diamine. We considered water-stabilized carbamic acid, but found it to be energetically unfavorable compared to ammonium carbamate.

    The binding energies are calculated in a systematic way, in order to replicate a likely local atomic environment that the adsorbing CO$_2$ would ``see''. In particular, we build up the binding energies of an adsorbed CO$_2$ based on the structure with already adsorbed CO$_2$, so that the binding of the $i^{\text{th}}$ CO$_2$ molecule is computed as:

    \begin{equation}\label{eq:binding-e}
        \Delta E_{\text{CO$_2$}:i+1} = E_{\text{MOF+CO$_2$}:i+1} - E_{\text{MOF+CO$_2$}:i} - E_{\text{CO$_2$}},
    \end{equation}

    \noindent where this difference is calculated for 0, 1, 2, and 3 H$_2$O per adsorption site.

    Based on the considerations of H$_2$O binding outlined in Section \ref{subsec:structure-gen}, and to try and understand the \textit{local} effect of H$_2$O on CO$_2$ adsorption, we restricted the H$_2$O locations to only be near the `lane' of amines in which CO$_2$ was adsorbing. This approach is physically justifiable, especially as we work towards a lattice model (discussed later in the paper), which can have 1, 2, or 6 lanes. Considering long-range effects, like the ones just discussed, is more suited to the 6 lane model, and is ripe for exploration in future studies. The focus of this paper is a 1 lane model (Fig. \ref{fig:struct_overview}(d)).

    \begin{figure}
        \centering
        \includegraphics[width=\linewidth]{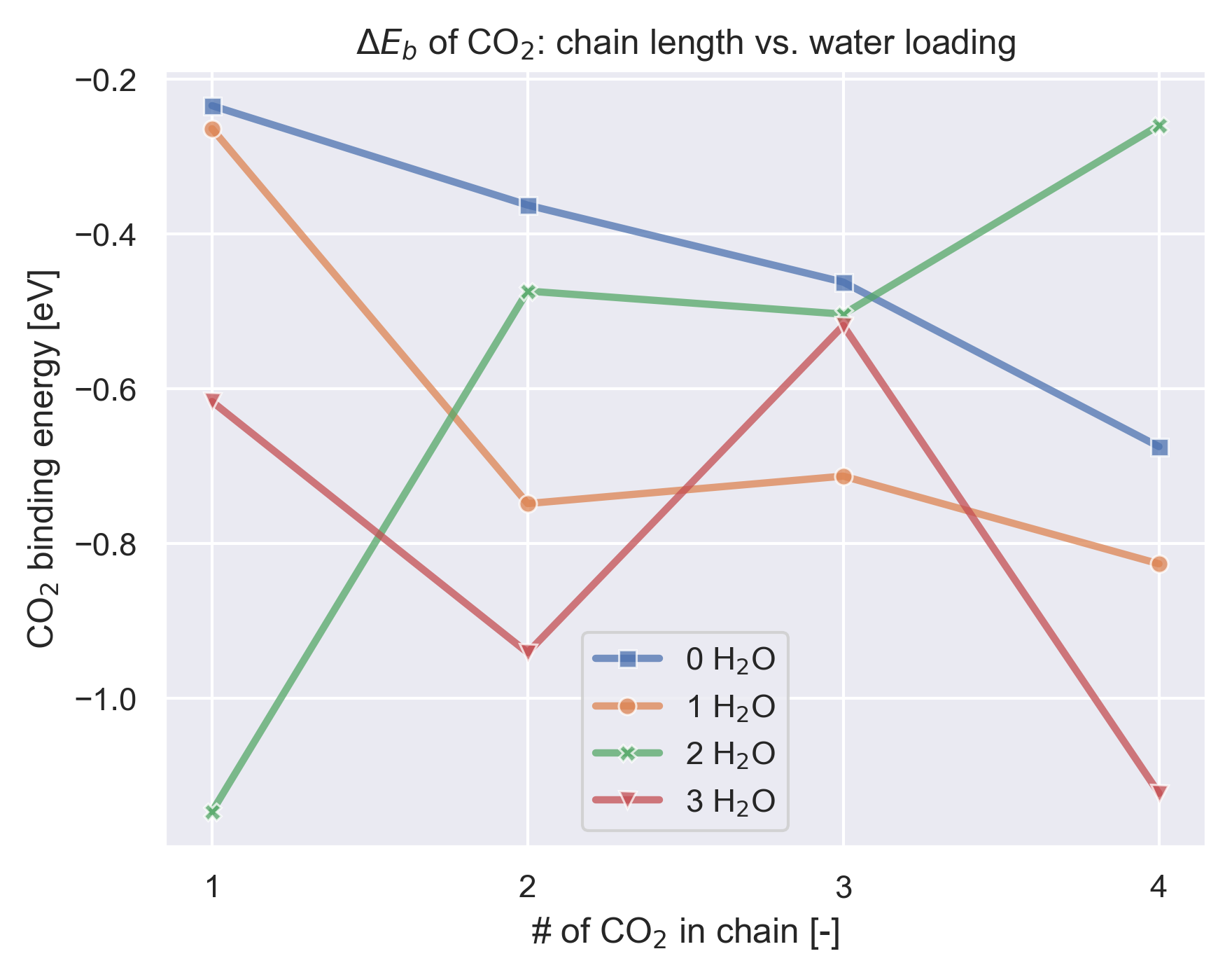}
        \caption{The energetics of CO$_2$ adsorption of chain forming for different amounts of water: 1, 2, or 3 moles of H$_2$O per diamine in the interacting lane. The blue curve is reproduced from Kundu, \textit{et al.}\cite{kundu_cooperative_2018}}
        \label{fig:chain-length-and-h2o}
    \end{figure}

    Given the identified binding locations, we can get a picture of the energetics of CO$_2$ adsorption for various chain lengths, with and without H$_2$O, as plotted in Fig. \ref{fig:chain-length-and-h2o}. In the absence of water, chains of longer length become more favorable.

    The addition of 1 H$_2$O per diamine makes the CO$_2$ binding energies more favorable, but doesn't change the qualitative nature of the behavior. The addition of another H$_2$O molecule per diamine, the green line in Fig. \ref{fig:chain-length-and-h2o}, completely reverses the chaining favorability, wherein the first insertion adsorption is very favorable, and subsequent adsorptions are much less so. In this case, then, CO$_2$ prefers to find isolated sites for the insertion-based adsorption. Lastly, when a third water molecule per diamine is added (the red line in Fig. \ref{fig:chain-length-and-h2o}), we see a hazier picture, wherein favorability moves around as the chain length grows.

    \begin{figure*}
        \centering
        \includegraphics[width=\linewidth]{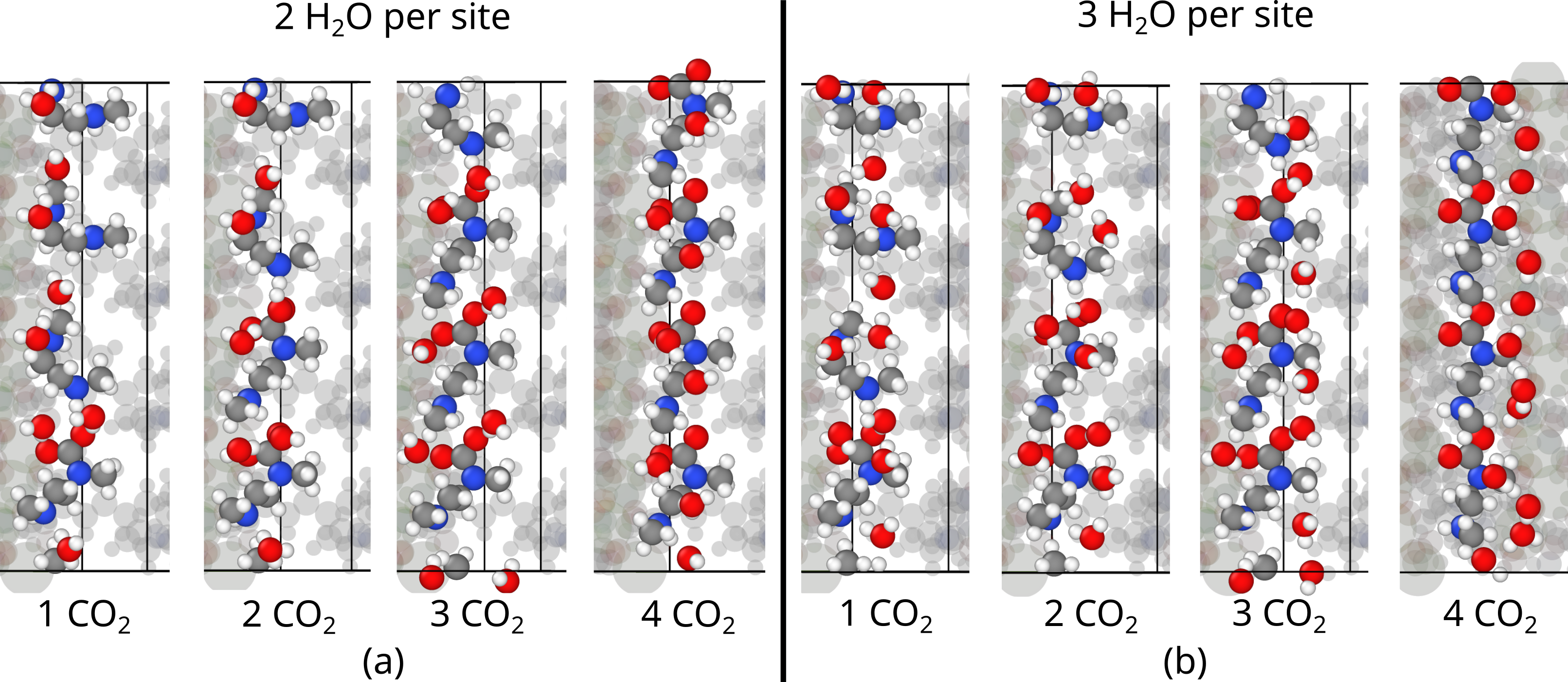}
        \caption{Structural changes upon increased number of adsorbed CO$_2$ molecules in a chain, in the case of (a) 2 and (b) 3 H$_2$O per adsorption site.}
        \label{fig:2and3co2_chaining}
    \end{figure*}

    Inspection of the atomic scale changes, with varying numbers of adsorbed water and CO$_2$ molecules, gives some physical insight into what is driving the energetics of adsorption. The cases of 2 and 3 H$_2$O molecules per adsorption site are shown in Figure \ref{fig:2and3co2_chaining}. The case of 1 H$_2$O per site was omitted, due to the energetic similarity of the chain build up to the dry case and relatively unremarkable structural changes.

    When there are 2 H$_2$O molecules at each lattice site (Fig. \ref{fig:2and3co2_chaining} (a)), structural changes that account for the extremely negative binding energy are induced upon the first CO$_2$ adsorption. Essentially, the CO$_2$ is not only forming its ammonium carbamate bond, but the movement of the amine makes the other amine group more amenable for H-bonding with a water molecule. The other H$_2$O molecules H-bonds to the CO$_2$, and a little clustering emerges. Inducing multiple bonds here explains the strong energetic favorability. Notice when there is no CO$_2$, the 2 H$_2$O molecules form a small pair. However, when a second CO$_2$ molecule is adsorbed in the chain, one of the H$_2$O molecules in the site below is pushed out of the way by the rotated methyl group of the amine that just adsorbed the CO$_2$, reducing some of the energetic favorability. This undoes some of the bonds that contributed to the very negative binding energy upon adsorption of the first CO$_2$. Adsorption at sequential sites breaks the H$_2$O pairs in the same way, by the rotation of the methyl group pushing one of the H$_2$O molecules into the pore.

    The case of 3 H$_2$O molecules per diamine is shown in Fig. \ref{fig:isothermsandchainlength}(b), building up from 1 to 4 CO$_2$ (an infinite chain). When the first CO$_2$ inserts itself, the waters local to the adsorption site participate in the insertion, slightly disrupting their local clustering but forming hydrogen bonds with the inserted CO$_2$. The second CO$_2$ insertion affects primarily its local H$_2$O, bumping them into a cluster, while still including them in hydrogen-bonding with the ammonium carbamate species. The addition of the third CO$_2$ seems to be something of an energy barrier (though still favorable), as we see some disruption of the local water clusters. Finally, when the final CO$_2$ adsorbs, we see an \textit{additional chain} emerge, that ties the water molecules up and down the lattice together. More interestingly, this chain is intertwined with and enabled by the CO$_2$ chain, introducing a braided CO$_2$/H$_2$O chain down the crystallographic \textit{c}-axis. This also supports previous work on a different amine showed that adsorption of CO$_2$ makes H$_2$O adsorption more favorable.

    \subsection{Lattice model}

    \subsubsection{Isotherms}\label{subsubsec:isotherms}

    To get a better appreciation of how the underlying physics manifests in the adsorption isotherm, we turn to an exactly-solvable statistical mechanical model.

    Kundu, \textit{et al.} show the uptake can be understood \textit{via} the adsorption energetics and the accessible free volume in three distinct adsorption configurations: a singly-adsorbed CO$_2$ molecule (1), an adsorbed CO$_2$ internal to a chain (int), and an adsorbed CO$_2$ forming the end of a chain (end), visualized in Fig. \ref{fig:struct_overview}(d). Mathematically, this is expressed as:

    \begin{equation}\label{eq:K-alpha}
        K_{\alpha} = \beta P V_{\alpha}e^{-\beta E_{\alpha}} \frac{q_{\text{inter}, \alpha}}{q_{\text{inter, bulk}}},
    \end{equation}

    \noindent where $\alpha = \{1, \text{int}, \text{end}\}$. $V_{\alpha}$ is the interaction volume in configuration $\alpha$, $E_{\alpha}$ is the binding energy of that configuration, $P$ is the pressure, $\beta = \frac{1}{k_B T}$ is the thermodynamic quantity, and $\frac{q_{\text{inter}, \alpha}}{q_{\text{inter, bulk}}}$ is the ratio of the vibrational and rotational partition functions of CO$_2$ in $\alpha$ and the bulk. The lattice model can be solved using transfer-matrix methods, and it can be shown the free energy is $f = -k_B T \text{ln}\lambda_+$, with $\lambda_+$ being the largest eigenvalue of the transfer matrix:

    \begin{equation}\label{eq:lambda_pos}
        2\lambda_+ = 1 + K_1 + K_{\text{int}} + \sqrt{(1 + K_1 - K_{\text{int}})^2 + 4K_{\text{end}}^2},
    \end{equation}

    The lattice site occupancies are given by

    \begin{equation}
        \rho = -\beta P (\partial f / \partial P).
    \end{equation}

    \noindent When multiplied by the maximum theoretical capacity ($q_{\infty} = 4.04$ mmol/g in the case of mmen-Mg$_2$(dobpdc)), governed by the chemistry of 1 CO$_2$ molecule per 2 amine groups, we get an uptake in mmol / g that is comparable with the experimentally measured isotherms.

    We want to adapt this model to incorporate water, while retaining its exact solvability and physical interpretability. As such, we made the following assumptions/decisions when extending the formalism: First, we are restricting our analysis to a single-lane model. This is related to the discussion earlier in the Section \ref{subsec:structure-gen}, in which we talk about the complications and variable nature of a full-pore of H$_2$O. A full pore of water would effectively break the single-lane approximation, as the large network of water would connect all six lanes in the structure. While this would be a more faithful representation of the likely ``true'' atomic-scale system, the addition of a large number of additional energy states would make an analytic solution intractable and more difficult to interpret, but is a direction for future study.

    Second, we are studying a scenario in which the lattice sites have already adsorbed the water molecules. In principal, this is reasonable, as the physisorptive nature of H$_2$O adsorption typically has faster kinetics over the chemisorptive nature of CO$_2$ adsorption. This could be rigorously implemented in a testing process, by first adsorbing H$_2$O to equilibrium at a given relative humidity, and then adsorbing CO$_2$. Additionally, as the presence of H$_2$O seems to enable the adsorption of CO$_2$ at previously inaccessible partial pressures\cite{siegelman_water_2019}, it seems likely the sites with H$_2$O already present will be the ones to adsorb the CO$_2$.

    Operating under these assumptions, we can treat the effect of H$_2$O as a modifier of the adsorption energetics for a given state ($E_\alpha$) and the corresponding accessible volume ($V_\alpha$). These changes will in turn affect the $K_{\alpha}$'s (eq. (\ref{eq:K-alpha})), and thus the isotherms.

    We make a few phenomenological modifications of the model, based on physical reasoning. The numbers we use as inputs to the lattice model are given in Table \ref{tab:binding_energies}. First, we note that the physical state corresponding to $K_1$ is a single inserted CO$_2$ molecule, a true seed point for chain formation. This is in contrast to the dry case, in which the carbamic acid adsorption via the non-bonded amine was most favorable. Given that, we modify the accessible volume of the first inserted CO$_2$ to be much smaller than in the carbamic acid case, as the insertion reaction allows much less movement. We don't lower it all the way to 11 \AA$^3$, however, because it can still wiggle, given that it has a free end, and the amine above it isn't locked in place. In the case of the internal chain site, we lower the accessible volume compared to the dry case, based on the observation H$_2$O molecules stay near the site, and lower the freedom of rotation of the amine with the adsorbed CO$_2$. Lastly, the end point has slightly higher freedom, because it is only bound on one side, but the other side is part of the tightly locked chain.

    It is crucial to stress here that these arguments are somewhat heuristic, and the step location is extremely sensitive to accessible free volume, with some uptakes at essentially $P = 0$ being $q_{\infty}$. This idea is explored in the SI of \textcite{kundu_cooperative_2018}. In the more recent, kinetics based work, the different physical parameters are taken from a fit of the model\cite{marshall_cooperative_2024}. In that case, the fit values for the accessible volume are approximately $10^{-6}$ \AA$^{3}$. The volumes in the original work were later corrected in an erratum\cite{kundu_erratum_2023}, wherein, due to a unit conversion error of the pressures, the reported volumes were actually five orders of magnitude smaller than what was originally reported. After correcting the pressure units, if the volumes are not modified to be on the order of $10^{-5}$, the cooperative adsorption step is at the completely wrong pressure compared to experiment. This can cause some difficulty in predicting the correct values for the $V_{\alpha}$, given that the geometrically estimated values are of such different orders of magnitude. However, the dimensionless ratio of the rotational and vibration partition functions for CO$_2$ in the state $\alpha$ and the bulk, can provide a scaling factor, wherein the base numbers for the volumes can be estimated geometrically, and a heuristic value of this ratio can fix the order of magnitude so that the model can predict isotherms in better agreement with the experiment\cite{kundu_cooperative_2018, marshall_cluster_2022, darunte_moving_2019}.

    In general, choosing the appropriate value of $V_{\alpha}$ from first principles is a challenge, as even small variations within the same order of magnitude can affect the step position\cite{kundu_cooperative_2018}. Tighter bounds on these values can be placed with experimental fits, but a full accounting for the sensitivity and a robust method of accurately estimating these quantities, particularly for a range of different diamine-appended MOFs, would be a promising direction for future studies. On the other hand, as the value of $V_{\alpha}$ don't appreciably affect the qualitative nature of the isotherms and adsorption behavior, our observations on the nature of H$_2$O's effect on the CO$_2$ adsorption remain unchanged.

    \begin{table}
        \centering
        \begin{tabular}{|c|c|c|c|c|c|c|}
            \hline
            \# H$_2$O & $E_1$ & $E_{\text{int}}$ & $E_{\text{end}}$ & $V_1$ & $V_{\text{int}}$ & $V_{\text{end}}$\\
            & [eV] & [eV] & [eV] & [\AA$^3$] & [\AA$^3$] & [\AA$^3$]
            \\\hline\hline
            0 & -0.23 & -0.72 & -0.57 & 500 & 11 & 11\\
            \hline
            1 & -0.26 & -0.82 & -0.73 & 20 & 5 & 7 \\
            \hline
            2 & -1.14 & -0.26 & -0.49 & 20 & 5 & 7 \\
            \hline
            3 & -0.61 & -1.12 & -0.73 & 20 & 5 & 7 \\
            \hline
        \end{tabular}
        \caption{Input parameters to the lattice model from \textit{ab initio} calculations. $E_1, E_{\text{int}}, E_{\text{end}}$ are respectively the adsorption energy of the first adsorbed CO$_2$, the adsorption energy of a CO$_2$ in the middle of a chain, and the adsorption energy of a CO$_2$ molecule that ends a chain. The $V$ values are analogous, but represent instead the accessible free volume of the adsorbed molecule in that configuration. Values from the first line are taken from \textcite{kundu_cooperative_2018}. The reported values for the $V_{\alpha}$ are scaled by $10^{-5}$ in the model, as discussed in the main text.}
        \label{tab:binding_energies}
    \end{table}

    The values in Table \ref{tab:binding_energies} are related to the values in Figure \ref{fig:chain-length-and-h2o} in the following way: The values for $E_1$ in all cases were directly taken for single CO$_2$ in the chain. For $E_{\text{end}}$, we take the average of the binding energy when there are 2 or 3 CO$_2$ in the chain. We tested the case when we took the binding energy for a chain of length 3, instead of the average, and the results were unaffected. For $E_{\text{int}}$, we take the value of the binding energy when there are 4 adsorbed CO$_2$, as this represents the bulk case, since our supercell has 4 \textit{c}-repetitions.

    \begin{figure*}
        \centering
        \includegraphics[width=\linewidth]{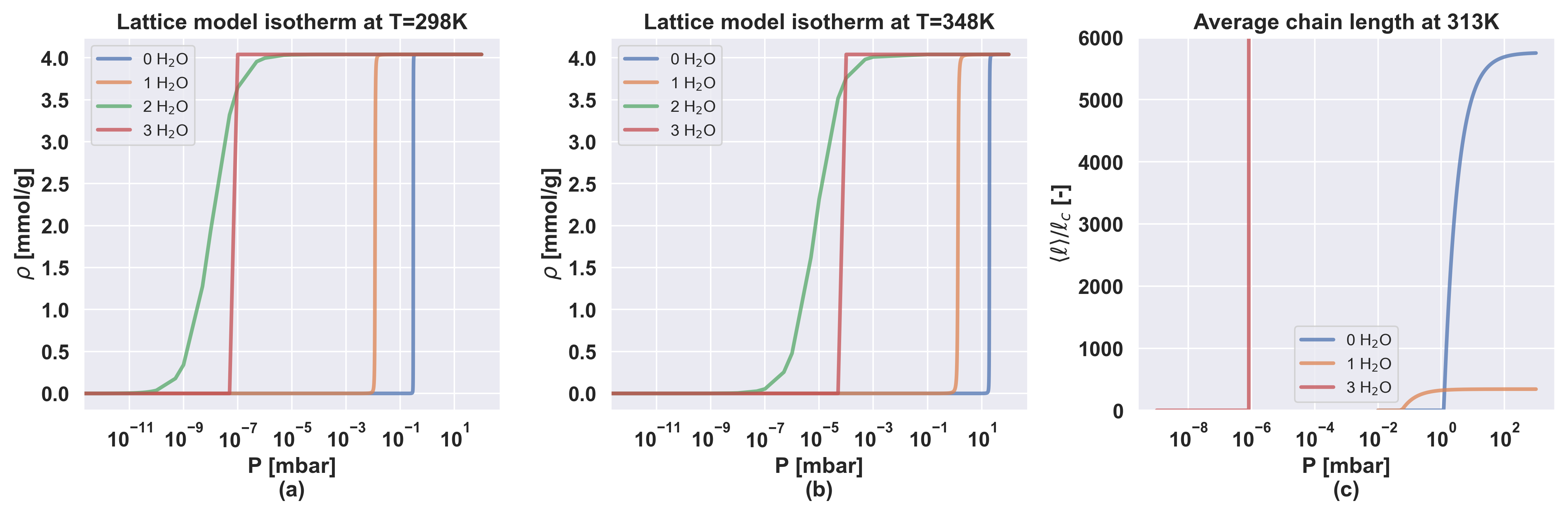}
        \caption{Single-lane lattice model predicted isotherms at (a) $T=298K$ and (b) $T=348K$ for the case of 0, 1, 2, and 3 H$_2$O molecules per diamine. The step location and shape of the isotherm depend heavily on the amount of water present. The step locations in the dry case agree with previously reported values in the literature, as does the phenomenon of increasing temperature shifting the step pressure to the right\cite{kundu_cooperative_2018,marshall_cluster_2022,darunte_moving_2019}. (c) Lattice model-predicted average chain length for the dry case (similar to what was previously reported in the literature\cite{kundu_cooperative_2018}) and the case of 1 H$_2$O per diamine. The lengths are scaled by $\ell_0 = 7$\AA, the approximate length of the Mg$_2$(dobpdc) unit cell in the \textit{c}-direction. In the cases of 2 H$_2$O molecules per diamine, the average chain length basically vanishes, as discussed in the main text of Section \ref{subsubsec:chainlengths}. When there are 3 H$_2$O per diamine, the average chain length diverges to infinity, behavior consistent with a phase transition.}
        \label{fig:isothermsandchainlength}
    \end{figure*}

    Inspection of Figure \ref{fig:isothermsandchainlength} shows how the isotherm is changed in the presence of varying levels of H$_2$O  at two temperatures. The dry cases (blue curves) correspond to the data reported in \textcite{kundu_cooperative_2018}. When a single H$_2$O molecule is present at a binding site, we effectively see the step simply shifted to a lower pressure. This makes sense, given that the binding energy in the 1 H$_2$O case is energetically more favorable, across the board. This is in agreement with previous studies, which argue that more efficient CO$_2$ uptake in the presence of H$_2$O is expressed in the increased favorability of the energetics of adsorption\cite{siegelman_water_2019}.

    When 2 H$_2$O molecules are present at each binding site, we can make some interesting observations. First, the shape of the isotherm is changed. It less resembles a step and more resembles a sigmoid function. Mathematically, this can be understood by the fact that single insertion binding is \textit{very favorable}. The pressures at which the uptake starts to increase are also \textit{extremely} small: $10^{-10}$ mbar. This approaches the realm of effectively exhibiting Langmuir-like adsorption behavior.

    Finally, when 3 H$_2$O molecules are present at a binding site, the step is almost a vertical line. From a numerical perspective, this is because the binding energy in all three situations is very favorable: a single molecule being adsorbed, a chain endpoint, and an internal chain member. The internal member is the strongest by a good margin, which explains the very steep slope. However, the fact that any situation in which the CO$_2$ can bind becomes favorable, means it makes sense that the uptake happens so rapidly.

    Increasing the temperature shifts all of the isotherms to the right, as previously reported and expected for these materials\cite{kundu_cooperative_2018}. The dry case continues to agree reasonably well with experiment. We expect that, if experimental data becomes available for this system, an increase in temperature would have a similar effect on the humid adsorption, as predicted here.

    \subsubsection{Average chain lengths}\label{subsubsec:chainlengths}

    In addition to the isotherms, one can use the lattice model to obtain quantitative information about chain lengths. Details of the derivations can be found in Kundu, \textit{et al.}\cite{kundu_cooperative_2018}. The governing equation for the average chain length is given by

    \begin{equation}
        \langle\ell\rangle = 2 + \frac{\rho_{\text{end}}\ell_0^2}{\rho_{\text{end}} + \rho_{\text{int}}},
    \end{equation}

    \noindent where $\ell_0 = -1 / \ln[2\rho_{\text{int}}/(\rho_{\text{end}} + 2\rho_{\text{int}})]$; $\rho_{\text{end}} = 2K^2_{\text{end}}\omega(1-\omega K_{\text{int}})/D$; $\rho_{\text{int}} = K^2_{\text{end}}\omega(\omega K_{\text{int}})/D$; $D = (1 + K_1)(1 - \omega K_{\text{int}})^2 + K^2_{\text{end}}\omega(2 - \omega K_{\text{int}})$; and $\omega = 1 / \lambda_+$. The $K_{\alpha}$ are as defined in eq. (\ref{eq:K-alpha}) and $\lambda_+$ is as defined in eq. (\ref{eq:lambda_pos}).

    The predicted average chain lengths are plotted in Figure \ref{fig:isothermsandchainlength}(c) for the 0 and 1 H$_2$O per diamine cases. The dry case agrees qualitatively with the previously reported plots. We expect some quantitative disagreement due to pressure conversion errors in the original paper\cite{kundu_erratum_2023}. The curves are qualitatively the same, and we see that a sharp increase in chain length occur at the step pressure, as expected, for the dry case. A similar phenomenon is observed in the case of 1 H$_2$O per diamine, though at a lower pressure, which agrees with the isotherm data in Section \ref{subsubsec:isotherms}. This increase is much less sharp than the dry case, however.

    In the case of 2 H$_2$O per diamine, the quantity essentially vanishes, and so we do not include it in the plot. This can be understand by looking at the equations in the limiting cased, coupled with the numerical values for the $K_{\alpha}$, as shown in Table \ref{tab:k_alpha}. The very favorable CO$_2$ binding energy for the first insertion in the presence of 2 H$_2$O molecules per diamine (-1.14 eV) means the $K_1$ term dominates the statistics. Mathematically speaking, we note that if $K_1 \gg K_{\text{int}, \text{end}}$, then the leading eigenvalue $\lambda_+$ (eq. (\ref{eq:lambda_pos})) becomes

    \begin{equation}
        \begin{split}
            \lambda_{+,2H_2O} &= 1 + K_1 + K_{\text{int}} + \sqrt{(1 + K_1 - K_{\text{int}})^2 + 4K_{\text{end}}^2} \\
            &\approx \frac{1}{2}(1 + K_1 + \sqrt{K_1^2}) \approx K_1.
        \end{split}
    \end{equation}

    \noindent This means that $\omega \approx 1 / K_{1,2H_2O}$ and that

    \begin{equation}
        \begin{split}
            D &= (1 + K_1)(1 - \omega K_{\text{int}})^2 + K^2_{\text{end}}\omega(2 - \omega K_{\text{int}}) \\
            &\approx (1 + K_1)(1 - K_{\text{int}}/K_{1,2H_2O})^2 \\
            &+ K^2_{\text{end}}/K_{1,2H_2O}(2 - K_{\text{int}}/K_{1,2H_2O}) \\
            &\approx K_{1,2H_2O}.
        \end{split}
    \end{equation}

    \noindent Given that $D \approx K_{1,2H_2O}$ is in the denominator of the expressions for $\rho_{\text{int,end}}$, we can conclude that $\rho_{\text{int,end}} \approx 0$ in the case of 2 H$_2$O per diamine, and thus the average chain length is negligible. This aligns with the sigmoid like shape observed in the isotherm, as seen in Figure \ref{fig:isothermsandchainlength} and discussed in Section \ref{subsubsec:isotherms}.

    A similar type of asymptotic analysis can be performed for the $K_{\alpha,3H_2O}$. In this case, $\lambda_+ \approx K_{\text{int}} \Rightarrow \omega = 1 / K_{\text{int}}$. That means that $D \approx 0$. In contrast to the 2 H$_2$O per diamine case, this means that we are likely to observe poor behavior in $\rho_{\text{int,end}}$ and $\ell_0$, as $D$ is in the denominator of these quantities, and thus $\langle\ell\rangle$ will also be poorly behaved. Indeed, we see that $\langle\ell\rangle$ diverges for the case of 3 H$_2$O per diamine, as shown in Figure \ref{fig:isothermsandchainlength}(c). Equivalently, this is like saying the average chain has \textit{infinite} length, and one could argue this prediction is in line with a phase transition\cite{kundu_cooperative_2018}.

    \begin{table}
        \centering
        \begin{tabular}{|c|c|c|c|}
            \hline
            \# H$_2$O & $K_1$ & $K_{\text{int}}$ & $K_{\text{end}}$ \\
            & [-] & [-] & [-]
            \\\hline\hline
            0 & $4.7\times 10^{-13}$ & $7.0\times 10^{-1}$ & $1.7\times 10^{-5}$ \\
            \hline
            1 & $6.9\times 10^{-15}$ & $3.1\times 10^{2}$ & $1.0\times 10^{-1}$ \\
            \hline
            2 & $1.1\times 10^{14}$ & $2.9\times 10^{-16}$ & $1.3\times 10^{-8}$\\
            \hline
            3 & $1.0\times 10^{-3}$ & $1.5\times 10^{12}$ & $1.3\times 10^{-7}$ \\
            \hline
        \end{tabular}
        \caption{ The $K_{\alpha}$ values for $T=313$K when there are 0, 1, 2, or 3 H$_2$O molecules per diamine, as computed with Equation \ref{eq:K-alpha}.}
        \label{tab:k_alpha}
    \end{table}

    \section{Discussion}

    On the whole, the results of the energetics of chain formation and the adapted lattice model paint a clear and feasible picture of the atomic scale behavior driving increased CO$_2$ adsorption in the presence of H$_2$O, both agreeing with and extending previous studies in the space. The cooperative nature of chain-formation in these materials is enhanced in the presence of a sufficient amount of H$_2$O, wherein it can form an additional and intertwined adsorbed chain with the MOF, CO$_2$, and H$_2$O. Strikingly, when a sufficient amount of H$_2$O is present, the predicted average chain length by our model tends to infinity, behavior more consistent with a phase transition.

    From a structural perspective, one of the dangling methyl groups in mmen plays a direct role in the energetics of the chain forming (Section \ref{subsec:chain-energetics}). This implies that the behavior is likely amine species dependent. It is well-established in the literature studying dry CO$_2$ adsorption in a variety of diamine- and tetramine-functionalized materials that the structure of the amine molecule affects the isotherm step location and shape\cite{siegelman_controlling_2017}. For example, ampd-Mg$_2$(dobpdc) displays a 2-step isotherm\cite{marshall_cooperative_2024, siegelman_water_2019}, as does the tetramine spermine\cite{kim_cooperative_2020}. A larger-scale study, exploring the same physical quantities (chain forming energetics, atomic-scale water/CO$_2$ interaction, \textit{etc.}) on a wider array of materials from the literature would give further understanding to why the different amines have the isotherms they do.

    Owing to the previously-discussed sensitivities of the model to changes in the binding energies and accessible volumes, the extremely low partial pressures displaying high uptake in Fig. \ref{fig:isothermsandchainlength} may be more qualitative than quantitative. However, the model's predictions could provide an additional interpretation of similar studies\cite{siegelman_water_2019,marshall_cooperative_2024}. In that work, they report the change of isotherm from a Type V to the Type I -- if uptake begins at partial pressures as low as those predicted here, it could present similar to a Langmuir isotherm.

    Generalization to a 2- or 6-lane model should account for more physical effects, particularly small amounts of pre- and post-step adsorption, something mostly absent from the 1-lane model. The challenge, of course, is that the long-range effects of H$_2$O need to be captured in the energy states of the lattice model, which would require a different model than the one employed here. Based on what was originally reported\cite{kundu_cooperative_2018}, we don't believe that the inclusion of the additional lanes will \textit{qualitatively} alter the details of the isotherms, but will instead change the fine features. Especially given that (with the exception of the 2 H$_2$O per diamine case), the binding energy of a CO$_2$ internal to the chain is much more energetically favorable than dimer binding\cite{kundu_cooperative_2018,marshall_cooperative_2024}, we expect the energetics of chain formation to still dominate the statistics of the model. Even in the case of 2 H$_2$O per diamine, the binding of a single CO$_2$ is much more favored, and the qualitative change in the shape of the isotherm from a step to more Langmuir like would manifest in a similar way if the presence of water drove more ``independent adsorption'' in adjacent lanes. Lastly, we note that, based on the localized nature of the chaining mechanism for the H$_2$O loading levels we studied, there don’t seem to be enough H$_2$O molecules at these relative humidities to cause inter-lane interaction. Even still, a thorough investigation of these effects will be a fruitful avenue of future work.

    There is no experimental data, as far as the authors are aware, of H$_2$O/CO$_2$ co-adsorption of mmen-Mg$_2$(dobpdc), against which to compare our predictions. We chose to study this system (as opposed to ampd-Mg$_2$(dobpdc), which has some co-adsorption data) because of the generalization of the results in the prior work by Kundu, \textit{et al.}\cite{kundu_cooperative_2018}, as well as this material's suitability to the 1-lane model.

    \section{Conclusion}

    In this work, we have combined \textit{ab initio} calculations and a statistical mechanical single-lane lattice model to study the co-adsorption of CO$_2$ and H$_2$O in an amine-functionalized MOF. In incrementally building up the ammonium carbamate chains, while varying the amount of H$_2$O, we were able to paint a picture of how the energetics of CO$_2$ adsorption behave in different scenarios. The observations add support the findings of prior studies, while the atomic-scale nature of the work elucidates the local behavior driving the material performance. We further predict an intriguing possible mechanism, in which the CO$_2$ and H$_2$O molecules combine to form an intertwined chain up and down the pore-axis of the material. The isotherm and chain length analysis from the lattice model in the case of 3 H$_2$O molecules per diamine, where this intertwined chain emerges, predict infinite average chain lengths, potentially pointing a change from cooperative capture to a true phase transition.

    Future studies will focus on extending the model to the full 6-lane picture, as well as allowing for a more nuanced study of the available energy states for H$_2$O in the partition function (such as bicarbonate), ideally expanding to a larger number of different amines. Another future avenue of investigation is understanding how the metal species affects the predicted and observed behavior. Ni, for example, does not have an experimental step, but its possible that H$_2$O could change the relative energetic favorability of the binding states to induce a step. Magnetic behavior and spin configurations could complicate these considerations\cite{jodaeeasl_comprehensive_2025}.

    \section*{Data Availability Statement}

    The data that supports the findings of this study are available within the article and its supplementary material.

    \section*{Supplementary Material}

    The SI includes the relaxed structures for all H$_2$O and CO$_2$ loadings studied, as well as tabular data of the energies of each system used in computing the binding energies.

    \begin{acknowledgments}
        This research used resources of the National Energy Research Scientific Computing Center (NERSC), a U.S. Department of Energy Office of Science User Facility located at Lawrence Berkeley National Laboratory, operated under Contract No. DE-AC02-05CH11231 using NERSC Award No. ALCC-ERCAP0025949. The authors would like to thank the many members of the carbon capture sorbent development team at GE Vernova Advanced Research, particularly Anil Duggal, for their insights and discussions.
    \end{acknowledgments}

    \bibliography{references_1}

\end{document}